\documentclass[peerreview, 11pt]{IEEEtran}
\usepackage{graphics}
\usepackage{epsfig}

\usepackage{subfigure}
\usepackage{graphics}

\usepackage{amsthm}
\usepackage{amsmath}
\usepackage{amsfonts}
\usepackage{mathtools}
\mathtoolsset{showonlyrefs=true}
\usepackage{dsfont}
\usepackage{amssymb}
\usepackage{color}
\definecolor{Blue}{rgb}{0.3,0.3,0.9}
\usepackage{stfloats}

\usepackage{stmaryrd}

\newcommand{\rem}[1]{}

\newtheorem{lemma}{Lemma}

\newtheorem{theorem}{Theorem}

\newcommand{\bre}{\begin{equation}}
\newcommand{\ere}{\end{equation}}

\newcommand{\ee}\]
\newcommand{\bra}{\begin{eqnarray}}
\newcommand{\era}{\end{eqnarray}}
\newcommand{\bfg}{\begin{figure}[hbtp]}
\newcommand{\efg}{\end{figure}}
\newcommand{\bver}{\begin{verbatim}}
\newcommand{\ever}{\end{verbatim}}
\newcommand{\bit}{\begin{itemize}}
\newcommand{\eit}{\end{itemize}}
\newcommand{\ben}{\begin{enumerate}}
\newcommand{\een}{\end{enumerate}}

\newcommand{\given}{\: | \:}

\newcommand{\tB}{\tilde B}

\newcommand{\baa}{\begin{eqnarray*}}
\newcommand{\eaa}{\end{eqnarray*}}

\newcommand{\bX}{{\bf X}}

\newcommand{\bY}{{\bf Y}}

\newcommand{\xor}{\oplus}
\newcommand{\bu}{{\bf u}}

\newcommand{\bx}{{\bf x}}
\newcommand{\by}{{\bf y}}

\newcommand{\cA}{{\cal A}}

\newcommand{\cE}{{\cal E}}

\newcommand{\cS}{{\cal S}}
\newcommand{\cT}{{\cal T}}
\newcommand{\cX}{{\cal X}}
\newcommand{\cY}{{\cal Y}}

\newcommand{\cU}{{\cal U}}

\newcommand{\cD}{\mathcal{D}}

\newcommand{\rE}{{\rm E}}

\def\defined{\: {\stackrel{\scriptscriptstyle \Delta}{=}} \: }

\def\defined{\: {\stackrel{\scriptscriptstyle \Delta}{=}} \: }

\newfont{\boldlarge}{msbm10 scaled 1100}

\newcommand{\comment}[1]{}

\newlength{\tmpbigbar}

\begin{document}

\title{Improved Bounds on the Finite Length Scaling of Polar Codes}

\author{Dina Goldin and David~Burshtein,~\IEEEmembership{Senior Member,~IEEE}
%\thanks{This research was supported by the Israel Science Foundation, grant no. 772/09.}
\thanks{D.\ Burshtein is with the school of Electrical Engineering, Tel-Aviv University, Tel-Aviv 69978, Israel (email: burstyn@eng.tau.ac.il).}
\thanks{D.\ Goldin is with the school of Electrical Engineering, Tel-Aviv University, Tel-Aviv 69978, Israel (email: dinagold@post.tau.ac.il).}
}

\markboth{Submitted to IEEE Transactions on Information Theory}{Goldin and Burshtein: Improved Bounds on the Finite Length Scaling of Polar Codes}

\maketitle \setcounter{page}{1}
%\IEEEpeerreviewmaketitle

\begin{abstract}
Improved bounds on the blocklength required to communicate over binary-input channels using polar codes, below some given error probability, are derived. For that purpose, an improved bound on the number of non-polarizing channels is obtained.
The main result is that the blocklength required to communicate reliably scales at most as $O((I(W)-R)^{-5.77})$ where $R$ is the code rate and $I(W)$ the symmetric capacity of the channel, $W$. The results are then extended to polar lossy source coding at rate $R$ of a source with symmetric distortion-rate function $D(\cdot)$.
The blocklength required scales at most as $O((D_N-D(R))^{-5.77})$ where $D_N$ is the actual distortion.
\end{abstract}

\begin{IEEEkeywords}
Channel polarization, polar codes, gap to capacity, rate distortion, finite length scaling.
\end{IEEEkeywords}

\section{Introduction} \label{sec:intro}
Polar codes, introduced by Arikan~\cite{arikan2009channel}, is an exciting recent development in coding theory. Arikan showed that, for a sufficiently large blocklength, polar codes can be used for reliable communications at rates arbitrarily close to the symmetric capacity (i.e., the mutual information between a uniform input distribution and the channel output) of an arbitrary binary-input channel.
Arikan also proposed encoding and decoding schemes, whose complexities scale as $O(N\log N)$ where $N$ is the blocklength of the code.
For $N$ sufficiently large, if the code rate  is below the symmetric capacity, then the error probability is upper bounded by $2^{-N^\beta}$ for any $\beta<1/2$~\cite{arikan2009channel},~\cite{arikan2009rate}. In~\cite{sasoglu2009polarization} it was shown that the results can be generalized for reliable communications below channel capacity over arbitrary discrete memoryless channels (DMCs). A very attractive property of polar codes is that although they are powerful, they are much simpler to analyze compared to low-density parity-check (LDPC) codes.
The main drawback of polar codes compared to LDPC-like codes is their inferior performance for codes with short to moderate blocklength size. However, recently it was shown~\cite{tal2011list} that the performance can be considerably improved by using a successive cancelation (SC) list decoder instead of the standard SC decoder, and by incorporating CRC bits.

Although originally proposed for channel coding, polar codes were extended to lossless and to lossy source coding~\cite{korada2010polar}, \cite{arikan2010source}.
In particular, Korada and Urbanke~\cite{korada2010polar} showed that for any design distortion, there exists a sequence of polar codes with
arbitrarily small redundancy, defined as the gap between the actual code rate and the rate-distortion function (evaluated at the design distortion). The encoding and decoding complexities are $O(N\log N)$.
Polar codes were also proposed to various other problems in multiuser information theory, including the Wyner-Ziv and Gelfand-Pinsker problems~\cite{korada2010polar}, write once memories (WOMs)~\cite{bur_str_full_version} and various other problems.

The rate of channel polarization was first studied in~\cite{arikan2009channel},~\cite{arikan2009rate}. Improved asymptotic upper bounds on the error rate that are code rate dependent, were presented in~\cite{tanaka2010refined}, \cite{tanaka2010speed}, \cite{hassani2010scaling1} and~\cite{hassani2013rate}. These results were obtained by analyzing the rate of convergence of the polarizing sub-channels.

All the results in the papers mentioned above assume that the blocklength, $N$, is sufficiently large, and do not specify how large it should be. The gap between the symmetric channel capacity and the polar code rate required for reliable communication, as a function of the blocklength, was discussed in~\cite{korada2010empirical}, \cite{goli2012universal} and~\cite{hassani2013finite}. A related result concerns the number of non-polarizing sub-channels for the binary erasure channel (BEC)~\cite{hassani2010scaling2}. In~\cite{goli2012universal}, \cite{hassani2013finite}, a binary memoryless symmetric (BMS) channel, $W$, with capacity $I(W)$ was considered (for BMS channels the symmetric capacity is also the capacity). Suppose that we use a polar code with blocklength $N$ and rate $R$ using the SC decoder, and that the error probability is bounded above by $P_e>0$. Also suppose that the sum of Bhattacharyya parameters is used as an approximation to the block error probability. Under this approximation, it was shown that we must have $N \ge \alpha / \left( I(W) - R \right)^{\tilde\mu}$. Here, $\alpha$ is a constant that depends only on $P_e$ and $R$, and the scaling parameter satisfies $\tilde{\mu} \ge 3.553$. It was further conjectured that the largest possible $\tilde{\mu}$ is $\tilde{\mu} = 3.627$, the parameter corresponding to the case where $W$ is a BEC.
In~\cite{guruswami2013speed}, \cite{hassani2013finite} it was further shown, under similar conditions but without the need to approximate the error probability by the sum of Bhattacharyya parameters, that it is sufficient to have $N = \beta / \left( I(W) - R \right)^\mu$ (or larger). Here $\beta$ is a constant that depends only on $P_e$ and $R$. The best scaling law was obtained in~\cite{hassani2013finite}, were it was shown that $\mu=7$ is sufficient. In this paper we improve this result to $\mu = 5.77$. We also extend the results to binary polar lossy source coding at rate $R$ of a source with symmetric distortion-rate function $D(\cdot)$. Denote the blocklength by $N$, the average distortion by $D_N$, and the redundancy by $\cD_N(R) \defined D_N - D(R)$. Then in order to obtain a redundancy at most $\cD^0$, it is sufficient to have $N = \beta / \left( \cD^0 \right)^{5.77}$ (or larger), where $\beta$ is a constant that depends only on $R$ and the properties of the source and the distortion measure used.

The rest of this paper is organized as follows. In Section~\ref{sec:background} we provide a brief background on polar codes. In Section~\ref{sec:cap_gap} we present our main results in this paper. First we derive an upper bound on the fraction of non-polarizing sub-channels. Then we obtain an upper bound on the blocklength required to communicate over a given binary-input channel, with error probability at most $P_e$, as a function of the gap between the symmetric capacity and the polar code rate. In Section~\ref{sec:extend} we extend our results to polar lossy source coding. Finally, Section~\ref{sec:discussion} concludes the paper.

\section{Background on Polar Codes}\label{sec:background}
We will follow the notation in~\cite{arikan2009channel}. Consider a binary-input discrete memoryless channel (B-DMC) $W:\cX\rightarrow\cY$ with input alphabet $\cX=\left\{0,1\right\}$ and output alphabet\footnote{The assumption that the channel is discrete is made for notational convenience only. For continuous output channels, sums should be replaced by integrals.} $\cY$. The \emph{symmetric capacity} of the channel, $I(W)$, is the mutual information between a uniform input distribution and the channel output, i.e.,\footnote{The base of all logarithms in this paper is $2$.}
\begin{equation}
I(W)=\sum_{x\in\cX}\sum_{y\in\cY}\frac{1}{2}W\left(y\given x\right) \log \frac{W\left(y\given x\right)} {\sum_{x'\in\cX}\frac{1}{2}W\left(y\given x'\right)}\;.
\label{eq:Idef}
\end{equation}
The Bhattacharyya parameter of the channel is defined by
\begin{equation}
Z(W)=\sum_{y\in\cY} \sqrt{W\left(y\given 0\right)W\left(y\given1\right)}\;.
\label{eq:Zdef}
\end{equation}

Let
$G_2 = \left( \begin{array}{cc}
1 & 0 \\
1 & 1 \\
\end{array} \right)$
and let its $n$th Kronecker product be $G_2^{\otimes n}$. Also denote $N=2^n$. Let $\bu = u_1^{N}$ be an $N$-dimensional binary $\{0,1\}$ message vector, and let $P_N$ be the bit-reversal permutation matrix, such that if $v_1^N = u_1^N P_N$ then $v_{b_1 \ldots b_n} = u_{b_n \ldots b_1}$ for all $b_1,\ldots,b_n \in \{0,1\}$. We can now define a generator matrix $G_N = P_N G_2^{\otimes n}$ and
$\bx = x_1^{N} = \bu G_N$ where the matrix multiplication is over ${\rm GF}(2)$. Suppose that we transmit $\bx$ over a B-DMC with transition probability $W(y \given x)$ and channel output vector $\by = y_1^{N}$. If $\bu$ is chosen at random with uniform probability, $1/2^N$, then the resulting probability distribution $P(\bu,\bx,\bu)$ is given by
\begin{equation}
\label{eq:p_uxy}
P(\bu,\bx,\by) = \frac{1}{2^N} \mathds{1}_{\{\bx = \bu G_N\}} \prod_{i=1}^{N} W(y_i \given x_i)
\end{equation}
Define the following $N$ sub-channels,
$$
W_N^{(i)}(\by,u_1^{i-1} \given u_i)
=
P(\by,u_1^{i-1} \given u_i)
=
\frac{1}{2^{N-1}} \sum_{u_{i+1}^{N}} P(\by \given \bu)
$$
Denote by $Z(W_N^{(i)})$ the Bhattacharyya parameters of the sub-channels $W_N^{(i)}$. In~\cite{arikan2009channel},~\cite{arikan2009rate} it was shown that asymptotically in $N$, a fraction $I(W)$ of the sub-channels satisfy $Z(W_N^{(i)}) < 2^{-N^\beta}$ for any $0 < \beta < 1/2$. Based on this result the following communication scheme was proposed. Let $R$ be the code rate. Denote by $F$ the set of $N(1-R)$ sub-channels with the highest values of $Z(W_N^{(i)})$ (the {\em frozen set}), and by $F^c$ the remaining $N \cdot R$ sub-channels. Fix the input to the sub-channels in $F$ to some arbitrary frozen vector $\bu_{F}$ (known both to the encoder and to the decoder) and use the channels in $F^c$ to transmit information. The encoder then transmits $\bx = \bu G_N$ over the channel. Recalling that $\bu_F$ is common knowledge, the decoder applies the following SC scheme.
For $i=1,2,\ldots,N$: If $i\in F$ then $\hat{u}_i = u_i$. Otherwise
\begin{equation}
\hat{u}_i = \left\{
              \begin{array}{ll}
                0 & \hbox{if $L_N^{(i)} > 1$} \\
                1 & \hbox{if $L_N^{(i)} \le 1$}
              \end{array}
            \right.
\label{eq:sc_decoder}
\end{equation}
where
\begin{equation}
L_N^{(i)} = L_N^{(i)}\left(\by,\hat{u}_1^{i-1}\right) = \frac{W_N^{(i)}(\by,\hat{u}_1^{i-1} \given u_i=0)}{W_N^{(i)}\left(\by,\hat{u}_1^{i-1} \given u_i=1\right)}
\label{eq:L_N}
\end{equation}
is the likelihood ratio of the channel $W_N^{(i)}(\by,\hat{u}_1^{i-1} \given u_i)$ corresponding to the channel output $\by,\hat{u}_1^{i-1}$. Asymptotically, reliable communication under SC decoding is possible for any $R<I(W)$. The error probability is upper bounded by $2^{-N^\beta}$ for any $\beta<1/2$,~\footnote{If the channel is BMS then this statement holds for any value of $\bu_F$. Otherwise, this statement is valid if $\bu_F$ is chosen uniformly at random.} and both the encoder and the SC decoder can be implemented in complexity $O(N\log N)$.

The analysis of polar codes is based on analyzing the evolution of the sub-channels $W_N^{(i)}$. The following recursion was obtained in~\cite{arikan2009channel}, for $N=2^n$, $n\ge 0$, and $i=1,\ldots,N$,
\begin{align}
\lefteqn{W_{2N}^{(2i-1)} \left( y_1^{2N},u_1^{2i-2} \given u_{2i-1} \right)} \nonumber \\
& & &= \frac{1}{2} \sum_{u_{2i}} W_{N}^{(i)} \left( y_1^{N},u_{1,o}^{2i-2} \xor u_{1,e}^{2i-2} \given u_{2i-1} \xor u_{2i} \right) W_{N}^{(i)} \left( y_{N+1}^{2N},u_{1,e}^{2i-2} \given u_{2i} \right) \label{eq:chan_recur1}\\
\lefteqn{W_{2N}^{(2i)} \left( y_1^{2N},u_1^{2i-1} \given u_{2i} \right)} \nonumber \\
& & &= \frac{1}{2} W_{N}^{(i)} \left( y_1^{N},u_{1,o}^{2i-2} \xor u_{1,e}^{2i-2} \given u_{2i-1} \xor u_{2i} \right) W_{N}^{(i)} \left( y_{N+1}^{2N},u_{1,e}^{2i-2} \given u_{2i} \right)
\label{eq:chan_recur2}
\end{align}
where $u_{1,o}^{2i-2}$ ($u_{1,e}^{2i-2}$, respectively) denote the odd (even) elements in the vector $u_1^{2i-2}$ . The recursion is initialized by $W^{(1)}_1=W$.

Following the notation in~\cite{korada2009polar} we now make the following additional definitions. Given two B-DMCs, $Q_1 \: : \: \cX \rightarrow \cY_1$ and $Q_2 \: : \: \cX \rightarrow \cY_2$, we define the following two channels, $Q_1 \boxast Q_2 \: : \: \cX \rightarrow \cY_1 \times \cY_2$ and $Q_1 \circledast Q_2 \: : \: \cX \rightarrow \cY_1 \times \cY_2 \times \cX$, by
\begin{align}
\left(Q_1 \boxast Q_2\right)\left(y_1,y_2 \given u\right) &\defined \frac{1}{2} \sum_x Q_1\left(y_1 \given u \xor x\right) Q_2\left(y_2 \given x\right) \label{eq:boxast_def}\\
\left(Q_1 \circledast Q_2\right)\left(y_1,y_2,x \given u\right) &\defined \frac{1}{2} Q_1\left(y_1 \given x \xor u\right) Q_2\left(y_2 \given u\right) \label{eq:circledast_def}
\end{align}
Using these definitions we can express~\eqref{eq:chan_recur1}-\eqref{eq:chan_recur2} as,
\begin{align*}
\tilde{W}_{2N}^{(2i-1)} \left( y_1^{2N}, u_{1,o}^{2i-2} \xor u_{1,e}^{2i-2}, u_{1,e}^{2i-2} \given u_{2i-1} \right) &=
(W_{N}^{(i)} \boxast W_{N}^{(i)})\left( y_1^{2N}, u_{1,o}^{2i-2} \xor u_{1,e}^{2i-2}, u_{1,e}^{2i-2} \given u_{2i-1} \right)\\
\tilde{W}_{2N}^{(2i)} \left( y_1^{2N}, u_{1,o}^{2i-2} \xor u_{1,e}^{2i-2},u_{1,e}^{2i-2},u_{2i-1} \given u_{2i} \right) &= (W_{N}^{(i)} \circledast W_{N}^{(i)})
\left( y_1^{2N}, u_{1,o}^{2i-2} \xor u_{1,e}^{2i-2},u_{1,e}^{2i-2},u_{2i-1} \given u_{2i} \right)
\end{align*}
Where $\tilde{W}_{2N}^{(2i-1)}$ and $\tilde{W}_{2N}^{(2i)}$ are the same channels as $W_{2N}^{(2i-1)}$ and $W_{2N}^{(2i)}$ (respectively) defined in~\eqref{eq:chan_recur1}-\eqref{eq:chan_recur2} up to a relabeling of the outputs. Hence, from an operational point of view, the channels $\tilde{W}_{2N}^{(2i-1)}$ and $\tilde{W}_{2N}^{(2i)}$ are identical to $W_{2N}^{(2i-1)}$ and $W_{2N}^{(2i)}$.

\begin{figure}
\hfill
\subfigure{\includegraphics[width=0.4\textwidth]{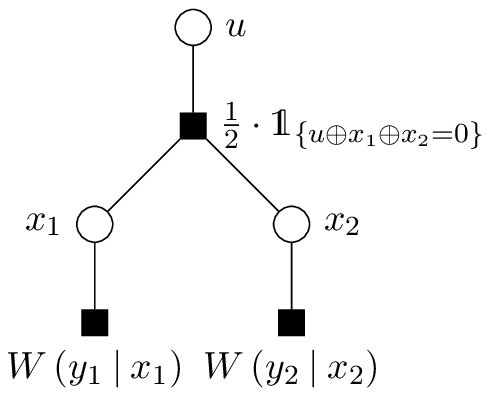}}
\hfill
\subfigure{\includegraphics[width=0.4\textwidth]{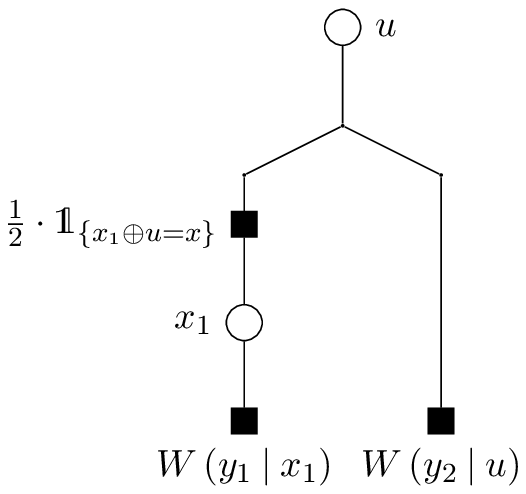}}
\hfill
\caption{The description of the $W^-$ (left) and $W^+$ (right) channels using factor graphs}
\label{W-+fig}
\end{figure}

By these observations we can now define a random process representing the evolution of the sub-channels as follows~\cite{arikan2009channel}, \cite{korada2009polar}. Let $B_1,B_2,\ldots$ be a sequence of independent identically distributed binary $\{0,1\}$ random variables that are uniformly distributed $\Pr \left\{ B_n = 0 \right\} = \Pr \left\{ B_n = 1 \right\} = 1/2$.
Let $W_0=W$ and let $W_n$ be defined recursively as follows,
\begin{equation}
W_{n+1} =
\left\{
  \begin{array}{ll}
    W_n^-, & \hbox{if $B_{n+1}=0$} \\
    W_n^+  & \hbox{if $B_{n+1}=1$.}
  \end{array}
\right.
\label{eq:Wn_recur}
\end{equation}
for $n=0,1,2,\ldots$ where
\begin{equation}
W^-\left(y_1,y_2 \given u\right) \defined (W \boxast W)(y_1,y_2 \given u) = \frac{1}{2} \sum_x W(y_1 \given u \oplus x) W(y_2 \given x)
\label{eq:W-def}
\end{equation}
and
\begin{equation}
W^+(y_1,y_2,x \given u) \defined (W \circledast W)(y_1,y_2,x \given u) = \frac{1}{2} W(y_1 \given x \oplus u) W(y_2 \given u)
\label{eq:W+def}
\end{equation}
The channels, $W^-$ and $W^+$, are depicted in Figure~\ref{W-+fig} using standard factor graph representations~\cite{ru_book}.

The random variable $W_n$ is uniformly distributed over the $N=2^n$ sub-channels $\left\{ W_N^{(i)} \right\}_{i=0}^{N-1}$. Hence, denoting by $I_n = I(W_n)$ and $Z_n = Z(W_n)$, we have,
$$
\Pr \left[ I_n \in (a,b) \right] = \left|\left\{ i \: : \: I\left(W_N^{(i)}\right) \in (a,b) \right\}\right| / \: N
$$
and
$$
\Pr \left[ Z_n \in (a,b) \right] = \left|\left\{ i \: : \: Z\left(W_N^{(i)}\right) \in (a,b) \right\}\right| / \: N
$$
Using these relations one can analyze the process~\eqref{eq:Wn_recur} and then obtain upper bounds on the error probability of polar codes under SC decoding~\cite{arikan2009channel}, \cite{arikan2009rate}.

It can be shown~\cite{khandekar2002graph}, \cite{arikan2009channel}, \cite{korada2009polar}, \cite{korada2010polar}, \cite{guruswami2013speed}, \cite{hassani2013finite} that
\begin{equation}
Z(W^+) = Z^2(W)
\label{eq:ZW+}
\end{equation}
and
\begin{equation}
Z(W) \sqrt{2-Z^2(W)} \le Z(W^-) \le 2 Z(W) - Z^2(W)
\label{eq:ZW-}
\end{equation}
The lower bound is obtained for a binary symmetric channel (BSC) and the upper bound for a BEC.

\section{Improved scaling results for polar channel codes} \label{sec:cap_gap}
Consider a binary-input channel $W$ with symmetric capacity $I(W)$. We wish to communicate over the channel using a polar code with blocklength $N$ and rate $R$. The error probability, when using the SC decoder, is required to be below $P_e$. In~\cite{guruswami2013speed}, \cite{hassani2013finite}, upper bounds on the required blocklength, $N$, were obtained under various conditions. In particular, in~\cite{hassani2013finite} it was shown that it is sufficient to have
\begin{equation}
\label{eq:hassani_result}
N = \beta / \left(I(W)-R\right)^{7}
\end{equation}
(or larger) where $\beta$ is a constant that depends only on $P_e$ and $I(W)$. The main tool in the proof was an upper bound on the number of non-polarizing sub-channels. In this section we first obtain an improved upper bound on the number of non-polarizing sub-channels. This bound will be used to obtain an improvement to~\eqref{eq:hassani_result}.

Given some function $f_0(z)$, defined over $[0,1]$ such that $f_0(z)>0$ for $z\in(0,1)$ and $f_0(0) = f_0(1) = 0$, we define $f_k(z)$ for $k=1,2,\ldots$ recursively as follows,
\begin{equation}
f_k(z)\triangleq\sup_{y\in\left[z\sqrt{2-z^2},z(2-z)\right]} \frac{f_{k-1}\left(z^2\right)+f_{k-1}(y)}{2} \label{f_k_def}
\;.
\end{equation}
It follows from this definition that for $k=0,1,2,\ldots$
\begin{equation}
f_k(0) = f_k(1) = 0
\label{eq:fk0_1}
\end{equation}
We also define
\begin{equation}
L_k(z) = \frac{f_k(z)}{f_0(z)} \quad , \quad L_k = \sup_{z\in(0,1)} L_k(z) \label{L_k_def}
\end{equation}
By these definitions we have, for $z\in (0,1)$
\begin{align}
f_k(z) &=
\sup_{y\in\left[z\sqrt{2-z^2},z(2-z)\right]} \frac{1}{2}
\left[ \frac{f_{k-1}(z^2)}{f_0(z^2)} f_0(z^2) + \frac{f_{k-1}(y)}{f_0(y)} f_0(y) \right]
\\ &\le
L_{k-1} \cdot \sup_{y\in\left[z\sqrt{2-z^2},z(2-z)\right]}
\frac{1}{2} \left[ f_0(z^2) + f_0(y) \right]
\\ &\le
L_{k-1} \cdot f_1(z)
\end{align}
Hence,
$$
\frac{f_k(z)}{f_0(z)} \le L_{k-1} \cdot \frac{f_1(z)}{f_0(z)} \le L_{k-1} \cdot L_1
$$
By taking the supremum of the left-hand side over $z\in (0,1)$ we obtain
$$
L_{k} \le L_{k-1} \cdot L_1
$$
Applying the last relation recursively $k$ times we thus conclude that
\begin{equation}
\label{eq:L_k_L_1}
\sqrt[k]{L_k} \le L_1
\end{equation}

\begin{lemma}\label{lemma:HassaniZk}
For every function $f_0(z)$ defined over $[0,1]$ such that $f_0(z)>0$ for $z\in(0,1)$ and $f_0(0)=f_0(1)=0$,
\begin{equation}
\label{eq:HassaniZk_1}
\rE\left[f_0\left(Z_n\right)\right]\le L_n \cdot f_0\left[Z(W)\right] \;.
\end{equation}
Furthermore, for any integer $0<k<n$,
\begin{equation}
\label{eq:HassaniZk_2}
\rE\left[f_0\left(Z_n\right)\right] \le
\left(\frac{L_1}{\sqrt[k]{L_k}}\right)^{k-1} \cdot
\left(\sqrt[k]{L_k}\right)^n \cdot f_0\left[Z(W)\right]
\end{equation}
\end{lemma}
For $k=1$, \eqref{eq:HassaniZk_2} degenerates to the result obtained in~\cite{hassani2013finite}.
Note that, due to~\eqref{eq:L_k_L_1}, the right-hand side of~\eqref{eq:HassaniZk_2} cannot have a slower rate of decay compared to the decay obtained in~\cite{hassani2013finite}.
\begin{IEEEproof}
First, we note that:
\begin{align}
\rE \left[f_k\left(Z_{n+1}\right)\right] &= \rE\left[\frac{f_k\left(Z_n^+\right)+ f_k\left(Z_n^-\right)}{2}\right]\\
&\le \rE\left[ \sup_{Z_n\sqrt{2-Z_n^2}\le y\le2Z_n-Z_n^2} \frac{f_k\left(Z_n^2\right)+ f_k\left(y\right)}{2} \right]\\
&= \rE\left[f_{k+1}\left(Z_n\right)\right]
\end{align}
where the inequality follows from~\eqref{eq:ZW+}--\eqref{eq:ZW-}, and the last equality follows from~\eqref{f_k_def}. Repeating this step $k$ times, we obtain
\begin{align}
\rE \left[f_0\left(Z_n\right)\right]&\le \rE\left[f_k\left(Z_{n-k}\right)\right]\\
&= \rE\left[\frac{f_k\left(Z_{n-k}\right)}{f_0\left(Z_{n-k}\right)} f_0\left(Z_{n-k}\right)\right]\le L_k \cdot \rE\left[f_0\left(Z_{n-k}\right)\right]\;.
\label{eq:Lk}
\end{align}
where the last inequality follows from \eqref{L_k_def}. Setting $k=n$ yields~\eqref{eq:HassaniZk_1}.
Now suppose that $n\equiv r\mod k$. We can reapply~\eqref{eq:Lk}, $\frac{n-r}{k}$ times, thus obtaining
\begin{align}
\rE\left[f_0\left(Z_n\right)\right]&\le L_k^{\frac{n-r}{k}} \rE \left[f_0\left(Z_r\right)\right]\\
&\le L_k^{\frac{n-r}{k}}\cdot L_1^r \cdot f_0\left[Z(W)\right]\\
&= \left(\frac{L_1}{\sqrt[k]{L_k}}\right)^{r}\cdot \left(\sqrt[k]{L_k}\right)^n \cdot f_0\left[Z(W)\right]
\end{align}
where the second inequality follows from~\eqref{eq:L_k_L_1} and~\eqref{eq:HassaniZk_1}. Now, using~\eqref{eq:L_k_L_1} and $r\le k-1$ we have,
$$
\left(\frac{L_1}{\sqrt[k]{L_k}}\right)^{r}\le \left(\frac{L_1}{\sqrt[k]{L_k}}\right)^{k-1}
$$
This yields~\eqref{eq:HassaniZk_2}.
\end{IEEEproof}

The computation of $L_k(z)$ for $z$ close to $0$ or $1$ can present numerical problems due to the division of zero by zero (see~\eqref{eq:fk0_1}). Consider the function $f_0(z)=z^\alpha (1-z)^{\beta}$, where $\alpha\in(0,1)$ and $\beta\in(0,1)$. As the following lemma shows, for this function we can calculate $L_k(z)$ analytically for $z$ close to zero or close to one. The main conclusion of the lemma is the last part.
\begin{lemma}
Suppose that $f_0(z) = z^\alpha (1-z)^{\beta}$ where $\alpha\in(0,1)$ and $\beta\in(0,1)$. Then,
\begin{enumerate}
\item \label{increasing0}
For each $k\ge 0$ there exists $0<\epsilon_k<1$ s.t. $f_k(z)$ is increasing in $(0,\epsilon_k)$. Furthermore,
\begin{equation}
\label{eq:fk1z}
f_{k+1}(z)=\frac{f_k\left(z^2\right)+f_k\left(2z-z^2\right)}{2} \quad z \in (0,\epsilon_{k+1})
\end{equation}
\item \label{decreasing1}
For each $k\ge 0$ there exists $0<\tilde\epsilon_k<1$ s.t. $f_k(z)$ is decreasing in $(\tilde\epsilon_k,1)$. Furthermore,
\begin{equation}
\label{eq:fk1za}
f_{k+1}(z)=\frac{f_k\left(z^2\right)+f_k\left(z\sqrt{2-z^2}\right)}{2} \quad z \in (\tilde\epsilon_{k+1},1)
\end{equation}
\item \label{taylor0}
For each $k \ge 0$ and finite $a$ and $b$
\begin{equation}
\lim_{z\rightarrow0^+}\frac{f_k\left[az+O\left(z^2\right)\right]}{f_k\left[bz+O\left(z^2\right)\right]}= \left(\frac{a}{b}\right)^\alpha
\end{equation}
\item \label{taylor1}
For each $k \ge 0$ and finite $a$ and $b$
\begin{equation}
\lim_{z\rightarrow1^-}\frac{f_k\left[1-a(1-z)+O\left[(1-z)^2\right]\right]} {f_k\left[1-b(1-z)+O\left[(1-z)^2\right]\right]}= \left(\frac{a}{b}\right)^\beta
\end{equation}
\item \label{main}
For each $k\ge 0$ and integer $m\ge 0$
\begin{align}
\lim_{z\rightarrow0^+}\frac{f_k\left(2^mz\right)}{2^mf_0(z)}&=\lim_{z\rightarrow0^+} \frac{f_0\left(2^{k+m}z\right)} {2^{k+m}f_0(z)}\\
\lim_{z\rightarrow1^-}\frac{f_k\left[1-2^m(1-z)\right]}{2^mf_0(z)}&=\lim_{z\rightarrow1^-} \frac{f_0\left[1-2^{k+m}(1-z)\right]} {2^{k+m}f_0(z)}
\end{align}
\item \label{main_main}
For each $k \ge 0$
\begin{align}
\lim_{z\rightarrow0^+}\frac{1}{k}\log\frac{f_k(z)}{f_0(z)}&= \alpha-1
\label{eq:Lz_k lim0}\\
\lim_{z\rightarrow1^-}\frac{1}{k}\log\frac{f_k(z)}{f_0(z)}&= \beta-1
\label{eq:Lz_k lim1}
\end{align}
\end{enumerate}
\label{lem:f_k(z) ratio}
\end{lemma}
\begin{IEEEproof}
The proof of part~\ref{increasing0}) follows by induction. The function $f_0(z)$ is indeed increasing for $z\in(0,\epsilon_0)$ for some $0<\epsilon_0<1$. We assume our claim is true for $k$, and prove it for $k+1$. Let $\epsilon_{k+1} \defined 1-\sqrt{1-\epsilon_k}$. Consider $z\in \left(0,\epsilon_{k+1}\right)$ (note that $1-\sqrt{1-\epsilon_k} \le \epsilon_k$). Then $z^2 \le 2z - z^2 \le \epsilon_k$.
Since $f_k(z)$ is increasing for $z \in (0,\epsilon_k)$, we obtain~\eqref{eq:fk1z} by the definition~\eqref{f_k_def}.
Furthermore, \eqref{eq:fk1z} shows that $f_{k+1}(z)$ is increasing for $z \in (0,\epsilon_{k+1})$ (for $z\in(0,\epsilon_{k+1})$, both $z^2$ and $2z-z^2$ are increasing and bounded above by $\epsilon_k$, and $f_k(z)$ is increasing for $z\in(0,\epsilon_k$)).

The proof of part~\ref{decreasing1}) is very similar and also follows by induction. The function $f_0(z)$ is indeed decreasing for $z\in(\tilde\epsilon_0,1)$ for some $0<\tilde\epsilon_0<1$. We assume our claim is true for $k$, and prove it for $k+1$. Let $\tilde\epsilon_{k+1} \defined \sqrt{\tilde\epsilon_k}$. Consider $z\in \left( \sqrt{\tilde\epsilon_k},1 \right)$ (note that $\sqrt{\tilde\epsilon_k} \ge \tilde\epsilon_k$). Then $z \sqrt{2-z^2} > z^2 > \tilde\epsilon_k$.
Since $f_k(z)$ is decreasing for $z \in (\tilde\epsilon_k,1)$, we obtain~\eqref{eq:fk1za} by the definition~\eqref{f_k_def}.
Furthermore, \eqref{eq:fk1za} shows that $f_{k+1}(z)$ is decreasing for $z \in (\tilde\epsilon_{k+1},1)$ (for $z\in(\tilde\epsilon_{k+1},1)$, both $z^2$ and $z \sqrt{2-z^2}$ are increasing and bounded below by $\tilde\epsilon_k$, and $f_k(z)$ is decreasing for $z\in(\tilde\epsilon_k$,1)).

The proof of part~\ref{taylor0}) follows by induction. Trivially, it is true for $k=0$. We assume our claim is true for $k$ and prove it for $k+1$.
\begin{align}
\lim_{z\rightarrow0^+}\frac{f_{k+1}\left[az+O\left(z^2\right)\right]} {f_{k+1}\left[bz+O\left(z^2\right)\right]}&=
\lim_{z\rightarrow0^+}\frac{f_k\left[O\left(z^2\right)\right]+f_k\left[2az+O\left(z^2\right)\right]} {f_k\left[O\left(z^2\right)\right]+f_k\left[2bz+O\left(z^2\right)\right]}\\
&=\lim_{z\rightarrow0^+}\frac{\frac{f_k\left[O\left(z^2\right)\right]}{f_k\left[2bz+O\left(z^2\right)\right]} +\frac{f_k\left[2az+O\left(z^2\right)\right]}{f_k\left[2bz+O\left(z^2\right)\right]}} {\frac{f_k\left[O\left(z^2\right)\right]}{f_k\left[2bz+O\left(z^2\right)\right]}+1}\\
&=\frac{0+\left(\frac{2a}{2b}\right)^\alpha}{0+1}=\left(\frac{a}{b}\right)^\alpha
\end{align}
where the first equality follows from~\eqref{eq:fk1z}, and the third equality follows from the induction assumption.

The proof of part~\ref{taylor1}) is very similar and also follows by induction. Trivially, it is true for $k=0$. We assume our claim is true for $k$, and prove it for $k+1$.
\begin{align}
\lim_{z\rightarrow1^-}\frac{f_{k+1}\left\{1-a(1-z)+O\left[(1-z)^2\right]\right\}} {f_{k+1}\left\{1-b(1-z)+O\left[(1-z)^2\right]\right\}}&=
\lim_{z\rightarrow1^-}\frac{f_k\left\{1-2a(1-z)+O\left[(1-z)^2\right]\right\}+f_k\left\{1+O\left[(1-z)^2\right]\right\}} {f_k\left\{1-2b(1-z)+O\left[(1-z)^2\right]\right\}+f_k\left\{1+O\left[(1-z)^2\right]\right\}}\\
&=\lim_{z\rightarrow1^-}\frac{\frac{f_k\left\{1-2a(1-z)+O\left[(1-z)^2\right]\right\}} {f_k\left\{1-2b(1-z)+O\left[(1-z)^2\right]\right\}} +\frac{f_k\left\{1+O\left[(1-z)^2\right]\right\}}{f_k\left\{1-2b(1-z)+O\left[(1-z)^2\right]\right\}}} {1+\frac{f_k\left\{1+O\left[(1-z)^2\right]\right\}}{f_k\left\{1-2b(1-z)+O\left[(1-z)^2\right]\right\}}}\\
&=\frac{\left(\frac{2a}{2b}\right)^\beta+0}{1+0}=\left(\frac{a}{b}\right)^\beta
\end{align}
where the third equality follows from the induction assumption, and the first equality follows from~\eqref{eq:fk1za} and the following relation for $z$ arbitrarily close to one,
\begin{multline}
\left\{1-a(1-z)+O\left[(1-z)^2\right]\right\}\sqrt{2-\left\{1-a(1-z)+O\left[(1-z)^2\right]\right\}^2}\\ =\left\{1-a(1-z)+O\left[(1-z)^2\right]\right\}\sqrt{1+2a(1-z)+O\left[(1-z)^2\right]}\\
=\left\{1-a(1-z)+O\left[(1-z)^2\right]\right\}\left\{1+a(1-z)+O\left[(1-z)^2\right]\right\}=1+O\left[(1-z)^2\right]\;. \label{relation}
\end{multline}

The proof of part~\ref{main}) also follows by induction. For $k=0$ the claim is trivial. Now we assume the statement is correct for $k$ and prove it for $k+1$. We have,
\begin{align}
\lim_{z\rightarrow0^+}\frac{f_{k+1}\left(2^mz\right)}{2^mf_0(z)}&= \lim_{z\rightarrow0^+}\frac{f_{k}(2^{m+1}z-4^m z^2)+f_k(4^mz^2)}{2^{m+1}f_0(z)}\\
&=\lim_{z\rightarrow0^+} \frac{f_{k}(2^{m+1}z)}{2^{m+1}f_0(z)}\\
&=\lim_{z\rightarrow0^+} \frac{f_{0}(2^{k+m+1}z)}{2^{k+m+1}f_0(z)}
\end{align}
where the first equality follows from \eqref{eq:fk1z}, the second follows from  part~\ref{taylor0}), and the third follows from the induction assumption.

Similarly, for $z\rightarrow 1^-$, the proof follows by induction. For $k=0$ the claim is trivial. Now we assume the statement is correct for $k$ and prove it for $k+1$. We have,
\begin{align}
\lim_{z\rightarrow1^-}\frac{f_{k+1}\left[1-2^m(1-z)\right]}{2^mf_0(z)}&=
\lim_{z\rightarrow1^-}\frac{f_{k}\left\{\left[1-2^m(1-z)\right]^2\right\}+ f_k\left\{\left[1-2^m(1-z)\right]\sqrt{2-\left[1-2^m(1-z)\right]^2}\right\}}{2^{m+1}f_0(z)}\\
&=\lim_{z\rightarrow1^-} \frac{f_{k}\left[1-2^{m+1}(1-z)\right]}{2^{m+1}f_0(z)}\\
&=\lim_{z\rightarrow1^-} \frac{f_{0}\left[1-2^{k+m+1}(1-z)\right]}{2^{k+m+1}f_0(z)}
\end{align}
where the first equality follows from \eqref{eq:fk1za}, the third follows from the induction assumption, and the second follows from part~\ref{taylor1}), using~\eqref{relation} with $a=2^m$.

Finally, the proof of part~\ref{main_main}) follows from part~\ref{main}) with $m=0$ as follows
\begin{align}
\lim_{z\rightarrow0^+}\frac{1}{k}\log\frac{f_k(z)}{f_0(z)}&= \lim_{z\rightarrow0^+}\frac{1}{k}\log \frac{f_0\left(2^kz\right)}{2^kf_0\left(z\right)}\\
&=\lim_{z\rightarrow0^+}\frac{1}{k}\log \frac{\left(2^kz\right)^\alpha}{2^kz^\alpha}= \alpha-1
\end{align}
and
\begin{align}
\lim_{z\rightarrow1^-}\frac{1}{k}\log\frac{f_k(z)}{f_0(z)}&= \lim_{z\rightarrow1^-}\frac{1}{k}\log \frac{f_0\left[1-2^k(1-z)\right]}{2^kf_0\left(z\right)}\\
&=\lim_{z\rightarrow1^-}\frac{1}{k}\log \frac{\left[2^k(1-z)\right]^\beta}{2^k(1-z)^\beta}= \beta-1
\end{align}
\end{IEEEproof}

\begin{figure}
\centering
\includegraphics[width=0.75\textwidth]{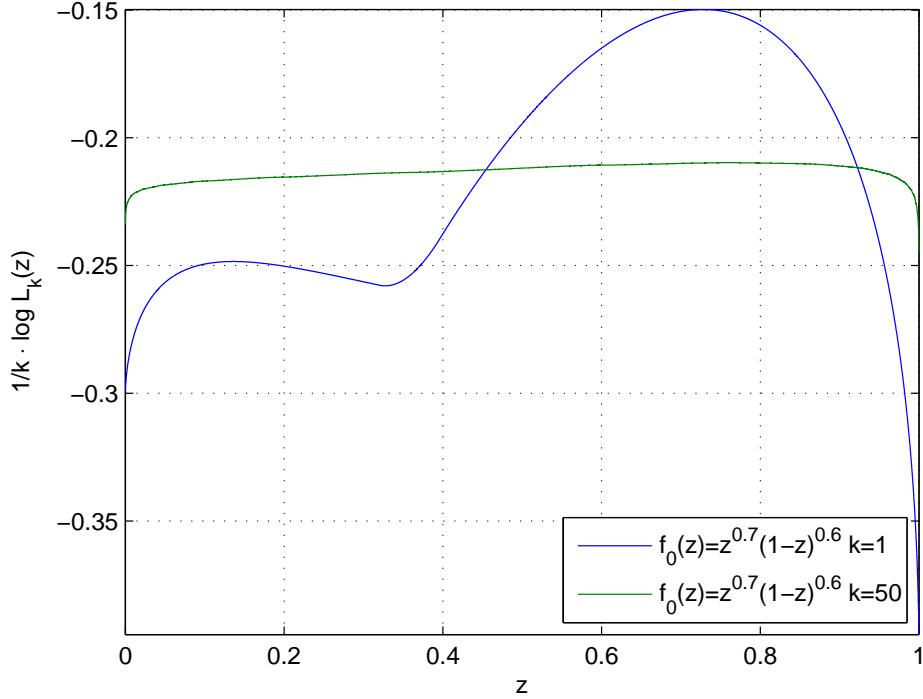}
\caption{A plot of $\frac{1}{k}\log L_k(z)$ for $k=1$ and $k=50$ when $f_0(z) = z^\alpha (1-z)^\beta$.} \label{L(z)_fig}
\end{figure}

Now suppose that $f_0(z)=z^{0.7}(1-z)^{0.6}$. As can be seen in Figure \ref{L(z)_fig}, we obtain $L_1=2^{-0.1498}$ and $\sqrt[50]{L_{50}}=2^{-0.2097}$.
Using~\eqref{eq:Lz_k lim0}--\eqref{eq:Lz_k lim1}, we see that for all values of $k$
\begin{align}
\lim_{z\rightarrow 0^+} \frac{1}{k} \log{L_k(z)} &= -0.3\\
\lim_{z\rightarrow 1^-} \frac{1}{k} \log{L_k(z)} &= -0.4
\end{align}
Note the sharp derivative of $f_k(z)$ for $z$ close to zero or one when $k$ is large.

\begin{figure}
\centering
\includegraphics[width=0.75\textwidth]{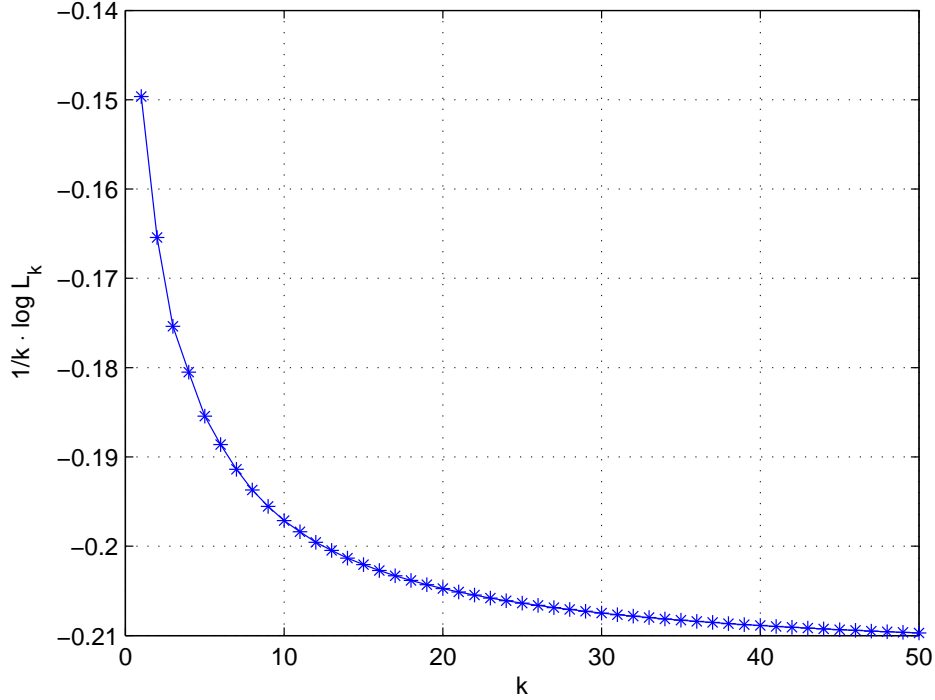}
\caption{A plot of $\frac{1}{k}\log {L_k}$ for $f_0(z)=z^{0.7}(1-z)^{0.6}$ as a function of $k$} \label{Lz(k)_fig}
\end{figure}

In Figure~\ref{Lz(k)_fig} we see that $\sqrt[k]{L_k}$ converges to a constant value for $k\rightarrow\infty$, and that it has almost converged for $k=50$.

Let $Y_n$ be defined by,
\begin{equation}
\label{eq:Y_n_def}
Y_n\defined\min\left(Z_n,1-Z_n\right)
\end{equation}
\begin{lemma} \label{lem:concave}
The following holds,
\begin{equation}
\Pr\left(Y_n>\delta\right)\le \frac{\alpha_1}{2\delta} \cdot 2^{-0.2097n}
\end{equation}
where $\alpha_1$ is some constant.
\end{lemma}
\begin{IEEEproof}
We use
$$
f_0(z)=z^{0.7}(1-z)^{0.6}
$$
It can be verified that $f_0(z)$ is concave. Combining this with $f_0(0)=f_0(1)=0$, we obtain $Y_n\le \frac{f_0\left(Z_n\right)}{2f_0(0.5)}$ (this inequality is verified for the two possible cases, $Z_n \le 1/2$ and $Z_n \ge 1/2$). Therefore, by Markov's inequality,
\begin{equation}
\Pr\left(Y_n>\delta\right) \le
\Pr\left( \frac{f_0(Z_n)}{2f_0(0.5)} > \delta \right) \le
\rE\left[f_0\left(Z_n\right)\right]\cdot\left(2f_0(0.5)\delta\right)^{-1} \label{markov-concave}
\end{equation}
Applying~\eqref{eq:HassaniZk_2} yields,
\begin{equation}
\Pr\left(Y_n>\delta\right)\le \left(\frac{f_0\left[Z(W)\right]}{2f_0(0.5)\delta}\right)\cdot \left(\frac{L_1}{\sqrt[k]{L_k}}\right)^{k-1} \left(\sqrt[k]{L_k}\right)^n
\end{equation}
Where $L_k$ is defined in~\eqref{L_k_def}.
As was noted above (Figure~\ref{L(z)_fig}), numerical calculations show that $L_1=2^{-0.1498}$, and for $k=50$, $\sqrt[k]{L_k}=2^{-0.2097}$. This proves our claim.
\end{IEEEproof}

We now need to translate this result on the rate of non-polarizing channels to a bound on the error rate. We could use the analysis of~\cite{hassani2013finite}. However, we present an alternative simple approach. This approach easily extends to the analysis of polar lossy source coding in the next section. We first state and prove the following.
\begin{lemma}\label{linkZ}
Suppose that
\begin{equation}
\label{eq:linkZ_assumption}
\Pr\left(\forall n\ge m_0\::\: Z_n \not\in (\delta,1-\delta) \right) \ge 1-\epsilon\\
\end{equation}
for some integer $m_0$, $0<\epsilon<1$ and $0<\delta<1/3$. Then
\begin{align}
\Pr\left(\forall n\ge m_0\::\: Z_n\le\delta\right)&\ge I(W)-\epsilon\\
\Pr\left(\forall n\ge m_0\::\: Z_n\ge 1-\delta\right)&\ge 1-I(W)-\epsilon
\end{align}
\end{lemma}
\begin{IEEEproof}
In~\cite{arikan2009channel} it was shown that $\lim_{n\rightarrow\infty}\Pr\left(Z_n\le\delta\right)=I(W)$ and $\lim_{n\rightarrow\infty}\Pr\left(Z_n\ge1-\delta\right)=1-I(W)$.
Assume
\begin{align}
\Pr\left(\forall n\ge m_0\::\: Z_n\le \delta\right)&=a_1\\
\Pr\left(\forall n\ge m_0\::\: Z_n\ge 1-\delta\right)&= a_2\;.
\end{align}
($a_1$ and $a_2$ depend on $m_0$). Combining this with~\eqref{eq:linkZ_assumption} yields
\begin{align}
1-\epsilon&\le\Pr\left(\forall n\ge m_0\::\: Z_n\notin(\delta,1-\delta)\right)\\
&=\Pr\left(\forall n\ge m_0\::\: Z_n\le\delta\right)+\Pr\left(\forall n\ge m_0\::\: Z_n\ge1-\delta\right)\\
&=a_1+a_2 \label{link*}
\end{align}
The first equality follows due to the assumption $\delta<1/3$ and~\eqref{eq:ZW+}--\eqref{eq:ZW-}, by which it follows that it is impossible to have $Z_n\le\delta$ and $Z_{n+1}\ge 1-\delta$ simultaneously, and it is also impossible that $Z_{n}\ge 1-\delta$ and $Z_{n+1}\le \delta$ simultaneously. That is,
$$
\{ \forall n\ge m_0\::\: Z_n\notin(\delta,1-\delta) \} =
\{ \forall n\ge m_0\::\: Z_n\le\delta \} \cup \{ \forall n \ge m_0\::\: Z_n \ge 1-\delta \}
$$
This explains the first equality in~\eqref{link*}. Now, clearly,
\begin{equation}
a_1 \le I(W) \quad , \quad a_2\le 1-I(W)\label{link**}
\end{equation}
We claim that $a_1\ge I(W)-\epsilon$ and $a_2\ge 1-I(W)-\epsilon$. By contradiction, assume that $a_1<I(W)-\epsilon$. Then, by \eqref{link**}, $a_1+a_2<1-\epsilon$, which contradicts \eqref{link*}. Therefore, $a_1\ge I(W)-\epsilon$. Similarly, assume by contradiction that $a_2<1-I(W)-\epsilon$. By \eqref{link**}, $a_1+a_2<1-\epsilon$, which contradicts \eqref{link*}. Therefore $a_2\ge 1-I(W)-\epsilon$.
\end{IEEEproof}

We can now state and prove our main result.
\begin{theorem}\label{theo:hassani17}
Suppose that we wish to use a polar code with rate $R$ and blocklength $N$ to transmit over a binary-input channel, $W$, with error probability at most $P_e>0$. Then it is sufficient to set
$$
N = \frac{\beta}{\left( I(W)-R \right)^{5.77}}
$$
(or larger) where $\beta$ is a constant that depends only on $P_e$.
\end{theorem}
Note: Our analysis can also be used to derive specific bounds on $N$ for a given value of $P_e$.

\begin{IEEEproof}
By Lemma~\ref{lem:concave} and Markov's inequality we have
\begin{equation}
\Pr\left(\exists n\ge m_0\::\:Z_n\in(\delta,1-\delta)\right)\le \sum_{n=m_0}^{\infty} \Pr \left( Y_n > \delta \right) \le \frac{\alpha_1}{2\delta} \cdot \frac{2^{-\rho m_0}}{1-2^{-\rho}}
\end{equation}
where $\rho = 0.2097$. That is,
\begin{equation}
\Pr\left(\forall n\ge m_0\::\: Z_n\notin(\delta,1-\delta)\right)\ge 1-\left(\frac{\alpha_1}{2\delta}\right)\cdot \frac{2^{-\rho m_0}}{1-2^{-\rho}}
\end{equation}
and together with Lemma \ref{linkZ} we obtain
\begin{align}
\Pr\left(\forall n\ge m_0\::\: Z_n\le\delta\right)&\ge I(W)-\left(\frac{\alpha_1}{2\delta}\right)\cdot \frac{2^{-\rho m_0}}{1-2^{-\rho}} \label{eq:Znzeta} \\
\Pr\left(\forall n\ge m_0\::\: Z_n\ge1-\delta\right)&\ge 1-I(W)-\left(\frac{\alpha_1}{2\delta}\right)\cdot \frac{2^{-\rho m_0}}{1-2^{-\rho}} \label{eq:Zn_ge_all_n}\;.
\end{align}

In~\cite[Section IV.B]{arikan2009channel}, Arikan defined the event
\begin{equation}
\cT_{m_0}(\delta)\triangleq\left\{Z_n\le\delta\quad\forall n\ge m_0\right\}
\end{equation}
Equation~\eqref{eq:Znzeta} can be rewritten as
\begin{equation}
\Pr\left[\cT_{m_0}(\delta)\right]\ge I(W)-\left(\frac{\alpha_1}{2\delta}\right)\cdot \frac{2^{-\rho m_0}}{1-2^{-\rho}}
\end{equation}
In~\cite[Section IV.B]{arikan2009channel}, Arikan also defined
\begin{equation}
\cU_{m_0,n}(\eta)\triangleq\left\{\sum_{i={m_0+1}}^{n}B_i>(0.5-\eta)(n-m_0)\right\} \label{calUdef}
\end{equation}
for $n>m_0\ge0$ and $0<\eta<0.5$.
In \cite[Equation (47)]{arikan2009channel} it was shown that
\begin{equation}
\Pr\left[\cU_{m_0,n}(\eta)\right]\ge 1-2^{-\left[1-h_2(0.5-\eta)\right](n-m_0)}\;.
\end{equation}
where $h_2(x) = -x\log x -(1-x)\log(1-x)$ is the binary entropy function. Applying the union bound yields
\begin{equation}
\Pr\left[\cT_{m_0}(\delta)\cap\cU_{m_0,n}(\eta)\right]\ge I(W)-2^{-\left[1-h_2(0.5-\eta)\right](n-m_0)}-\left(\frac{\alpha_1}{2\delta}\right)\cdot \frac{2^{-\rho m_0}}{1-2^{-\rho}}\;.
\end{equation}
If we pick
\begin{equation}
m_0=\frac{\left[1-h_2(0.5-\eta)\right]n}{1-h_2(0.5-\eta)+\rho} \label{rho*m_0=alpha*n}
\end{equation}
we obtain
\begin{equation}
\Pr\left[\cT_{m_0}(\delta)\cap\cU_{m_0,n}(\eta)\right]\ge I(W)-\left(1+\frac{\alpha_1} {2\delta\left(1-2^{-\rho}\right)}\right)\cdot 2^{-\alpha n}\;.
\end{equation}
where
\begin{equation}
\alpha=\left(\frac{1}{1-h_2(0.5-\eta)}+\frac{1}{\rho}\right)^{-1} \label{alpha_def}
\end{equation}
Clearly, $\lim_{\eta\rightarrow0.5}\alpha=\left(1+\frac{1}{\rho}\right)^{-1}$.
Arikan proved that if the event $\cT_{m_0}(\delta)\cap\cU_{m_0,n}(\eta)$ holds, then $Z_n\le\delta\left[2^{0.5+\eta}\delta^{0.5-\eta}\right]^{n-m_0}$. If we pick $m_0$ as in \eqref{rho*m_0=alpha*n} and
\begin{equation}
\log\delta=-\frac{1.5+\eta+\frac{1-h_2(0.5-\eta)}{\rho}}{0.5-\eta} -\kappa \label{log2zeta}
\end{equation}
where $\kappa>0$ is a constant, we obtain that if the event $\cT_{m_0}(\delta)\cap\cU_{m_0,n}(\eta)$ holds, then
\begin{equation}
Z_n\le\delta\cdot2^{-n\left[1+\frac{\kappa\rho(0.5-\eta)}{1-h_2(0.5-\eta)+\rho}\right]}
\end{equation}
Hence,
\begin{equation}
\Pr\left(Z_n\le\delta\cdot 2^{-n\left[1+\frac{\kappa\rho(0.5-\eta)}{1-h_2(0.5-\eta)+\rho}\right]} \right)\ge I(W)-\left(1+\frac{\alpha_1}{2\delta\left(1-2^{-\rho}\right)}\right)\cdot 2^{-\alpha n}
\end{equation}
Now, for every rate
\begin{equation}
R\le I(W)-\left(1+\frac{\alpha_1}{2\delta\left(1-2^{-\rho}\right)}\right)\cdot 2^{-\alpha n} \label{Rle}
\end{equation}
let $\cA_N$ be defined as the set of $N\cdot R$ smallest values of $\left\{ Z\left( W_N^{(i)} \right) \right\}_{i=1}^N$ ($\cA_N$ are the active channels, those that are not frozen). From the two inequalities above, we know that
\begin{equation}
\max_{i\in\cA_N} Z\left( W_N^{(i)} \right) \le \delta\cdot N^{-1-\frac{\kappa\rho(0.5-\eta)}{1-h_2(0.5-\eta)+\rho}}\;.
\end{equation}
From \cite[Proposition 2]{arikan2009channel} we know that $\Pr(\cE)\le\sum_{i\in\cA_N} Z\left( W_N^{(i)} \right)$. Putting this together, we obtain
\begin{equation}
\Pr(\cE) <
N R\max_{i\in\cA_N} Z\left( W_N^{(i)} \right) \le
R \delta \cdot N^{-\frac{\kappa\rho(0.5-\eta)}{1-h_2(0.5-\eta)+\rho}} \label{Pele}
\end{equation}

If we define $\Delta=I(W)-R$, then \eqref{Rle} becomes
\begin{equation}
\label{Rle1}
\log N\ge \left[\log\left(1+\frac{\alpha_1}{2\delta\left(1-2^{-\rho}\right)}\right)-\log \Delta \right] \left(\frac{1}{1-h_2(0.5-\eta)}+\frac{1}{\rho}\right)
\end{equation}
where $0<\eta<0.5$ and $0<\kappa$ are constants, and $\log\delta$ is defined in \eqref{log2zeta}.
In addition,
\begin{equation}
\log N\ge \left[\log\delta-\log P_e \right]\left(\frac{1-h_2(0.5-\eta)+\rho}{\kappa\rho(0.5-\eta)}\right)
\end{equation}
is equivalent to
\begin{equation}
\delta \cdot N^{-\frac{\kappa\rho(0.5-\eta)}{1-h_2(0.5-\eta)+\rho}}\le P_e\;.
\end{equation}
Since \eqref{Rle} (i.e.,~\eqref{Rle1}) yields~\eqref{Pele}, it follows that if
\begin{multline}
\log N\ge\max\left\{\left[\log\left(1+\frac{\alpha_1}{2\delta\left(1-2^{-\rho}\right)}\right)-\log \Delta \right] \left(\frac{1}{1-h_2(0.5-\eta)}+\frac{1}{\rho}\right)\right.,\\ \left.\left[\log\delta-\log P_e \right]\left(\frac{1-h_2(0.5-\eta)+\rho}{\kappa\rho(0.5-\eta)}\right)\right\}
\end{multline}
then
\begin{equation}
\Pr\left(\cE\right)\le R\delta \cdot N^{-\frac{\kappa\rho(0.5-\eta)}{1-h_2(0.5-\eta)+\rho}}\le P_e\;.
\end{equation}
We have thus obtained an upper bound on the blocklength required for communications with error probability at most $P_e$ as a function of the gap to the symmetric capacity.

Setting $\eta\rightarrow0.5^-$, (i.e., $\delta\rightarrow0^+$) and using the fact that, by Lemma~\ref{lem:concave}, $\rho=0.2097$ (so that $(1+1/\rho) < 5.77$) yields the required result.
\end{IEEEproof}

In appendix~\ref{app:IW} we briefly indicate how, instead of the Bhattacharyya parameter, we can use the symmetric capacity to derive bounds using a very similar approach.

\section{Scaling results for polar lossy source coding} \label{sec:extend}
\subsection{Background}
We start by providing a brief background on polar source coding~\cite{korada2010polar} (see also~\cite{karzand2010qsrc}, \cite{bur_str_full_version}). Consider some random variable $Y\in\cY$, and assume for simplicity that $\cY$ is finite. Also denote $\cX=\{0,1\}$. The source vector random variable, $\bY = Y_1^N$, is created by independent sampling of the source $Y$. Let $d(\bx,\by)$ be some finite distance measure between two $N$ dimensional vectors, $\bx=x_1^N$ and $\by=y_1^N$, such that $d(\bx,\by) = \sum_{i=1}^N d(x_i,y_i)$ where $d(x,y)$ is the distance between the symbols $x\in\cX$ and $y\in\cY$. Suppose that $d(x,y) \le d_{\max}$ for all $x\in\cX$ and $y\in\cY$ (in \cite[Lemma 5]{korada2010polar} $d_{\max}=1$). Given some distortion level, $D>0$, let $W(y \given x)$ be the test channel that achieves the symmetric rate-distortion, $R(D)$, of the source, defined as rate-distortion under the constraint that the input to the test channel, $X$, is uniformly distributed over $\cX$. A polar source code is then constructed using this test channel. The code has a frozen set $F$ that consists of the $(1-R) \cdot N$ sub-channels with the largest values of $Z\left(W_N^{(i)}\right)$. This code uses some arbitrary frozen vector $\bu_F$ which is known both to the encoder and to the decoder (e.g., $\bu_F=0$) and has rate $R = |F^c|/N$.
Given $\bY=\by$ the SC encoder applies the following scheme. For $i=1,2,\ldots,N$, if $i\in F$ then $\hat{u}_i = u_i$, otherwise
\begin{equation}
\label{eq:sc_encoder}
\hat{u}_i = \left\{
              \begin{array}{ll}
                0 & \hbox{w.p. $L_N^{(i)} / \left(L_N^{(i)} + 1\right)$} \\
                1 & \hbox{w.p. $1 / \left(L_N^{(i)} + 1\right)$}
              \end{array}
            \right.
\end{equation}
The complexity of this scheme is $O(N\log N)$.
Since $\hat{\bu}_F = \bu_F$ is common knowledge, the decoder only needs to obtain $\hat{\bu}_{F^c}$ from the encoder ($|F^c|$ bits). It can then reconstruct the approximating source codeword $\bx$ using $\bx = \hat{\bu} G_2^{\otimes n}$.
Let ${\rm E} d(\bX(\bY),\bY) / N$ be the average distortion  of this polar code (the averaging is over both the source vector, $\bY$, and over the approximating source codeword, $\bX(\bY)$, which is determined at random from $\bY$).
Denote by $D$ the design distortion (using which we construct the test channel and design the code), by $D_N = {\rm E} d(\bX(\bY),\bY) / N$ the actual distortion, and by $R$ the rate of the code.
In~\cite{korada2010polar} it was shown, for $N$ sufficiently large, that the rate, $R$, can approach the symmetric rate-distortion function, $R(D)$, arbitrarily close and at the same time
\bre
\label{eq:D_N_Avg}
D_N - D \le O\left(2^{-N^\beta}\right)
\ere
Note that if $W\left(y\given x\right)$ is a symmetric channel, the value of $\bu_F$ can be set arbitrarily. If $W\left(y\given x\right)$ is not symmetric, we must average over all $2^{\left|F\right|}$ choices of $\bu_F$ while calculating $D_N$ in order to obtain~\eqref{eq:D_N_Avg}.

\subsection{Upper Bound on the blocklength}
We now apply our results in Section~\ref{sec:cap_gap} to obtain upper bounds on the blocklength of polar lossy source codes.
\begin{theorem}\label{theo:hassani17S}
Suppose that we wish to use a polar code with rate $R$ for lossy source coding of some source with a symmetric distortion-rate function, $D(\cdot)$, with average distortion $D_N>D(R)$, redundancy $\cD_N(R)\defined D_N-D(R)$~\cite{zhang1997redundancy}, and blocklength $N$. Then, in order to obtain a redundancy at most $\cD^0$ (i.e., $\cD_N^0(R) \le \cD^0$) it is sufficient to set
$$
N = \frac{\beta}{\left( \cD^0 \right)^{5.77}}
$$
(or larger) where $\beta$ is a constant that depends only on $R$, $d_{\max}$ and $D(\cdot)$.
\end{theorem}

\begin{IEEEproof}
Denote by $D$ the design distortion, and by $I(W)$ the symmetric capacity of the test channel such that $I(W)=R(D)$~\cite{korada2010polar}.
We will follow the proof of Theorem~\ref{theo:hassani17}, replacing $Z_n$ with $1-Z_n^2$, as in the proof of~\cite[Theorem 19]{korada2010polar}.
If $Z_n\ge1-\delta$, then $1-Z_n^2\le2\delta-\delta^2\le2\delta$. Hence, by \eqref{eq:Zn_ge_all_n},
\begin{equation}
\Pr\left(\forall n\ge m_0\::\: 1-Z_n^2\le\delta\right) \ge 1-I(W)-\left(\frac{\alpha_1}{\delta}\right)\cdot \frac{2^{-\rho m_0}}{1-2^{-\rho}} \label{linkZsource}
\end{equation}
Define
\begin{equation}
\cS_{m_0}\left(\delta\right)\triangleq\left\{1-Z_n^2\le\delta\quad\forall n\ge m_0\right\}
\label{calSdef}
\end{equation}
for $\delta\ge0$ and $m_0\ge 0$. Rewriting~\eqref{linkZsource} we have,
\begin{equation}
\Pr\left[\cS_{m_0}(\delta)\right]\ge 1-I(W)-\left(\frac{\alpha_1}{\delta}\right)\cdot \frac{2^{-\rho m_0}}{1-2^{-\rho}}\;.
\end{equation}

In the proof of \cite[Theorem 19]{korada2010polar} it is shown that
\begin{align}
1-Z_{n+1}^2 \le
\left\{
  \begin{array}{ll}
    \left(1-Z_n^2\right)^2, & \hbox{if $B_{n+1}=0$;} \\
    2\left(1-Z_n^2\right), & \hbox{if $B_{n+1}=1$.}
  \end{array}
\right.
\end{align}
where the $B_n$ sequence was defined in Section~\ref{sec:background}. Hence, if the event $\cS_{m_0}(\delta)$ holds and $n\ge m_0$, then
\begin{equation}
\frac{1-Z_{n+1}^2}{1-Z_n^2}\le\left\{
\begin{array}{ll}
\delta & \hbox{if $B_{n+1}=0$;}\\
2 & \hbox{if $B_{n+1}=1$.}
\end{array}\right.
\end{equation}
(using~\eqref{eq:ZW+}--\eqref{eq:ZW-}).
Similarly to the proof of \cite[Theorem 2]{arikan2009channel}, if the event $\cS_{m_0}(\delta)$ holds and $n>m_0$, then
\begin{equation}
1-Z_n^2\le\delta\cdot2^{n-m_0}\cdot\prod_{i=m_0+1}^n\left(\delta/2\right)^{\tB_i}
\end{equation}
where $\tB_i=1-B_i$. Hence, if the event $\cS_{m_0}(\delta)\cap\tilde\cU_{m_0,n}(\eta)$ holds, then
\begin{equation}
1-Z_n^2\le\delta\cdot\left[2^{\frac{1}{2}+\eta}\delta^{\frac{1}{2}-\eta}\right]^{n-m_0}
\end{equation}
where the set $\tilde\cU_{m_0,n}(\eta)$ is defined as
\begin{equation}
\tilde\cU_{m_0,n}(\eta)\triangleq\left\{\sum_{i={m_0+1}}^{n}\tB_i>(0.5-\eta)(n-m_0)\right\}\;.
\end{equation}
Setting $m_0$ as in~\eqref{rho*m_0=alpha*n} and $\delta$ as in \eqref{log2zeta}, we obtain
\begin{equation}
1-Z_n^2\le\delta\cdot2^{-n\left[1+\frac{\kappa\rho(0.5-\eta)}{1-h_2(0.5-\eta)+\rho}\right]}
\end{equation}
if the event $\cS_{m_0}(\delta)\cap\tilde\cU_{m_0,n}(\eta)$ holds. Hence,
\begin{equation}
\Pr\left(1-Z_n^2\le\delta\cdot 2^{-n\left[1+\frac{\kappa\rho(0.5-\eta)}{1-h_2(0.5-\eta)+\rho}\right]} \right)\ge 1-I(W)-\left(1+\frac{\alpha_1}{\delta\left(1-2^{-\rho}\right)}\right)\cdot 2^{-\alpha n}
\end{equation}

For every rate
\begin{equation}
R\ge I(W)+\left(1+\frac{\alpha_1}{\delta\left(1-2^{-\rho}\right)}\right)\cdot 2^{-\alpha n} \label{Rge}
\end{equation}
we pick $F$ as the set of $N(1-R)$ largest values of $Z\left(W_N^{(i)}\right)$. Since $1-Z_n\le1-Z_n^2$, from the two inequalities above, we know that
\begin{equation}
\max_{i\in F}\left(1-Z\left(W_N^{(i)}\right)\right)\le \delta\cdot N^{-1-\frac{\kappa\rho(0.5-\eta)}{1-h_2(0.5-\eta)+\rho}}\;.
\end{equation}
From~\cite[Lemma 5]{korada2010polar} and~\cite[Lemma 7]{korada2010polar} we know that $D_N-D\le d_{\max}\sum_{i\in F}\sqrt{2\left(1-Z\left(W_N^{(i)}\right)\right)}$. Putting this together, we obtain
\begin{equation}
D_N-D\le d_{\max} N(1-R)\max_{i\in F}\sqrt{2\left(1-Z\left(W_N^{(i)}\right)\right)} \le d_{\max} \sqrt{2\delta}(1-R)N^{\frac{1}{2}-\frac{\kappa\rho(0.5-\eta)}{2\left[1-h_2(0.5-\eta)+\rho\right]}} \label{Dple}
\end{equation}
Defining $\Delta\defined R-I(W)$, we obtain
\begin{equation}
\cD_N(R)=D_N-D+D-D(R)=D_N-D+\Delta\left|\frac{D(R)-D}{R-I(W)}\right|\le D_N-D+\Delta\cdot\left|D'(I(W))\right|
\end{equation}
where $D'(x)=\frac{dD(x)}{dx}$. The last inequality follows from the convexity of $D(R)$. Note that in this bound we have one degree of freedom, the design distortion $D$, which defines the symmetric capacity $I(W)$ ($I(W)=R(D)$) of the test channel.
Setting $I(W)$ equal to the right-hand side in~\eqref{Rge} yields,
\begin{equation}
\label{eq:D_N(R)_bound}
\cD_N(R)\le d_{\max} \sqrt{2\delta}(1-R)N^{\frac{1}{2}-\frac{\kappa\rho(0.5-\eta)}{2\left[1-h_2(0.5-\eta)+\rho\right]}} +\left(1+\frac{\alpha_1}{\delta\left(1-2^{-\rho}\right)}\right)\cdot N^{-\left(\frac{1}{1-h_2(0.5-\eta)}+\frac{1}{\rho}\right)^{-1}}\cdot\left|D'(I(W))\right|\;.
\end{equation}

We now set $\eta\rightarrow0.5^-$ and $\kappa$ large so that $\delta\rightarrow0^+$. Furthermore, if $\kappa$ is sufficiently large then the second term in~\eqref{eq:D_N(R)_bound} is asymptotically dominant.
In addition, for $N = \beta/(\cD^0)^{5.77}$ where the constant $\beta$ is sufficiently large, we obtain
$$
I(W) = R - \left(1+\frac{\alpha_1}{\delta\left(1-2^{-\rho}\right)}\right)\cdot N^{-\alpha} > \frac{R}{2}
$$
Hence, due to convexity of $D(R)$, $\left|D'(I(W))\right| < \left|D'(R/2)\right|$. It follows from~\eqref{eq:D_N(R)_bound} (using $\rho=0.2097$ by Lemma~\ref{lem:concave}) that if $\beta$ is sufficiently large then $\cD_N(R) < \cD^0$.
\end{IEEEproof}

Zhang et al. proved in \cite{zhang1997redundancy}, that the best achievable distortion redundancy is $\cD(R)=\Theta\left(\frac{\ln N}{N}\right)$. Asymptotically, it is better than our results.

\section{Discussion} \label{sec:discussion}
In this paper we have considered a polar code with blocklength $N$ and rate $R$ transmitted over a binary-input channel, $W$, with symmetric capacity $I(W)$. Decoding is performed using the SC decoder. If the error probability needs to be below some $P_e>0$, then it is sufficient to have $N = \beta / \left( I(W) - R \right)^\mu$. Here $\beta$ is a constant that depends only on $P_e$, and $\mu = 5.77$. The results were also extended to polar source coding. The natural question to ask is what is the lowest possible value of the scaling parameter $\mu$. From the simulations presented in \cite[Figure 3]{hassani2013finite} it seems likely that we must have $\mu > 5$. Hence, the value of $\mu$ that we obtained seems close to the optimum. Nevertheless, further improvements in the bound on $\mu$ may perhaps be obtained. Our best results were obtained when using the Bhattacharyya parameter in the analysis. These results were better compared to the results obtained when using the symmetric capacity (i.e., mutual information) parameter. We have also made some efforts to work with the error rate and the channel parameter considered in~\cite{bur_bp_bounds}. These parameters were also inferior compared to the Bhattacharyya parameter. However, other channel parameters may possibly yield further improvements to our results.

The optimal scaling law of $N$ with respect to the gap to the symmetric capacity, $I(W)-R$, is $O(\left( I(W) - R \right)^{-2})$. Using polar codes we now know that the scaling law is $O\left(\left( I(W) - R \right)^{-\mu}\right)$ where $3.55 \le \mu \le 5.77$ (The lower bound, $3.55$, was obtained after approximating the block error probability by the sum of Bhattacharyya parameters, but it is also the scaling factor of the BEC). As noted in~\cite{hassani2013finite} the scaling can be improved by using more general polarization kernels. This topic is left for future research. Another possibility for future research concerns the blocklength scaling of nonbinary polar codes.

%\section*{Acknowledgment}
%The authors would like to thank ...

\appendices
\section{Scaling results using mutual information} \label{app:IW}
Assume for simplicity that the channel is BMS (in~\cite{korada2009polar} it is noted how to generalize to non-symmetric channels). It can be shown~\cite[Chapter 4]{ru_book} that the following inequalities hold,
\begin{align}
I(W^-) &\le 1 - h_2\left( 2h_2^{-1}(1-I(W))(1-h_2^{-1}(1-I(W))) \right) \label{eq:IW-max}\\
I(W^-) &\ge I^2(W) \label{eq:IW-min}
\end{align}
In addition,
\begin{equation}
I(W^+)+I(W^-)=2I(W)
\label{eq:sum_IW}
\end{equation}

Motivated by these inequalities, we modify the definition of $f_k(z)$ as follows. Given some function $f_0(x)$, defined over $[0,1]$ such that $f_0(x)>0$ for $x\in(0,1)$, and $f_0(0)=f_0(1)=0$, we define $f_k(x)$ for $k=1,2,\ldots$ recursively as follows,
\begin{equation}
f_k(x)\triangleq\sup_{\epsilon_l(x)\le \epsilon\le\epsilon_h(x)} \frac{f_{k-1}(x+\epsilon)+f_{k-1}(x-\epsilon)}{2}
\end{equation}
where $\epsilon_l(x)$ and $\epsilon_h(x)$ are defined by
\begin{align}
\epsilon_l\left( x \right) &=
x + h_2\left\{2h_2^{-1}\left[1-x\right]\cdot \left[1-h_2^{-1}\left[1-x\right]\right]\right\}-1 \label{epsl}\\
\epsilon_h\left(x\right) &= x - x^2\;. \label{epsh}
\end{align}

The definitions of $L_k(x)$ and $L_k$ are the same as in~\eqref{L_k_def}.
With the new definition of $f_k(x)$, Equation~\eqref{eq:L_k_L_1} still holds.
Similarly to~\eqref{eq:HassaniZk_2} we have, for an integer $0<k<n$,
\begin{equation}
\rE\left[f_0\left(I_n\right)\right] \le
\left(\frac{L_1}{\sqrt[k]{L_k}}\right)^{k-1} \cdot
\left(\sqrt[k]{L_k}\right)^n \cdot f_0\left[I(W)\right]
\end{equation}
Similarly to~\eqref{eq:Y_n_def} we define $J_n \defined \min(I_n,1-I_n)$. Using the concave function
$$
f_0(x)=\left(1-\sqrt{1-x^2}\right)^{0.402}\left(1-x^{1.11}\right)^{0.604}
$$
we obtain, similarly to Lemma~\ref{lem:concave},
\begin{equation}
\Pr\left(J_n>\delta\right)\le \frac{\alpha_1}{2\delta} \cdot 2^{-0.1786n}
\end{equation}
Numerical calculations yield $L_1=2^{-0.1708}$ and, for sufficiently large $k$, $\sqrt[k]{L_k} \le 2^{-0.1786}$.

Similarly to Lemma~\ref{linkZ} we have the following. If
\begin{equation}
\label{eq:linkI_assumption}
\Pr\left[\forall n\ge m_0\::\: I_n \not\in (\delta,1-\delta) \right] \ge 1-\epsilon\\
\end{equation}
for some integer $m_0$, $0<\epsilon<1$ and $\delta<1/3$. Then
\begin{align}
\Pr\left(\forall n\ge m_0\::\: I_n\ge1-\delta\right)&\ge I(W)-\epsilon\\
\Pr\left(\forall n\ge m_0\::\: I_n\le\delta\right)&\ge 1-I(W)-\epsilon
\end{align}
The proof is essentially the same as the proof of Lemma~\ref{linkZ}, with $I_n$ replacing $1-Z_n$.
Finally, we can obtain a result similar to Theorem~\ref{theo:hassani17}. We use essentially the same proof but with the following modification. First we obtain a result similar to~\eqref{eq:Znzeta} using the same approach:
$$
\Pr\left(\forall n\ge m_0\::\: I_n\ge 1-\delta\right)\ge I(W)-\left(\frac{\alpha_1}{2\delta}\right)\cdot \frac{2^{-\rho m_0}}{1-2^{-\rho}}
$$
Then we combine it with~\cite[Equation (2)]{arikan2009channel} to obtain,
$$
\Pr\left(\forall n\ge m_0\::\: Z_n\le\zeta\right)\ge I(W)-\left(\frac{\alpha_1}{\zeta^2}\right)\cdot \frac{2^{-\rho m_0}}{1-2^{-\rho}}
$$
and proceed with the derivation in Theorem~\ref{theo:hassani17}.

However, this time we can only claim that it is sufficient to set
$$
N = \frac{\beta}{\left( I(W)-R \right)^{6.6}}
$$
(or larger), where $\beta$ is a constant that depends only on $P_e$, since now $\rho=0.1786$.

%\bibliography{bibliography}
\bibliographystyle{IEEEtran}
\bibliography{IEEEabrv,bibliography}

\begin{thebibliography}{10}
\providecommand{\url}[1]{#1}
\csname url@samestyle\endcsname
\providecommand{\newblock}{\relax}
\providecommand{\bibinfo}[2]{#2}
\providecommand{\BIBentrySTDinterwordspacing}{\spaceskip=0pt\relax}
\providecommand{\BIBentryALTinterwordstretchfactor}{4}
\providecommand{\BIBentryALTinterwordspacing}{\spaceskip=\fontdimen2\font plus
\BIBentryALTinterwordstretchfactor\fontdimen3\font minus
  \fontdimen4\font\relax}
\providecommand{\BIBforeignlanguage}[2]{{%
\expandafter\ifx\csname l@#1\endcsname\relax
\typeout{** WARNING: IEEEtran.bst: No hyphenation pattern has been}%
\typeout{** loaded for the language `#1'. Using the pattern for}%
\typeout{** the default language instead.}%
\else
\language=\csname l@#1\endcsname
\fi
#2}}
\providecommand{\BIBdecl}{\relax}
\BIBdecl

\bibitem{arikan2009channel}
E.~Arikan, ``{Channel polarization: A method for constructing
  capacity-achieving codes for symmetric binary-input memoryless channels},''
  \emph{IEEE Transactions on Information Theory}, vol.~55, no.~7, pp.
  3051--3073, 2009.

\bibitem{arikan2009rate}
E.~Arikan and E.~Telatar, ``{On the rate of channel polarization},'' in
  \emph{Proc. IEEE International Symposium on Information Theory (ISIT)}, 2009,
  pp. 1493--1495.

\bibitem{sasoglu2009polarization}
E.~Sasoglu, E.~Telatar, and E.~Arikan, ``{Polarization for arbitrary discrete
  memoryless channels},'' in \emph{Proc. IEEE Information Theory Workshop
  (ITW)}, 2009, pp. 144--148.

\bibitem{tal2011list}
I.~Tal and A.~Vardy, ``{List decoding of polar codes},'' in \emph{Proc. IEEE
  International Symposium on Information Theory (ISIT)}, Saint Petersburg,
  Russia, August 2011, pp. 1--5.

\bibitem{korada2010polar}
S.~B. Korada and R.~L. Urbanke, ``{Polar codes are optimal for lossy source
  coding},'' \emph{IEEE Transactions on Information Theory}, vol.~56, no.~4,
  pp. 1751--1768, 2010.

\bibitem{arikan2010source}
E.~Arikan, ``{Source polarization},'' in \emph{Proc. IEEE International
  Symposium on Information Theory (ISIT)}, Austin, Texas, June 2010, pp.
  899--903.

\bibitem{bur_str_full_version}
D.~Burshtein and A.~Strugatski, ``{Polar write once memory codes},'' \emph{IEEE
  Transactions on Information Theory}, vol.~59, no.~8, pp. 5088--5101, August
  2013.

\bibitem{tanaka2010refined}
T.~Tanaka and R.~Mori, ``{Refined rate of channel polarization},'' in
  \emph{Proc. IEEE International Symposium on Information Theory (ISIT)},
  Austin, Texas, June 2010, pp. 889--893.

\bibitem{tanaka2010speed}
T.~Tanaka, ``{On speed of channel polarization},'' in \emph{Proc. IEEE
  Information Theory Workshop (ITW)}, Dublin, Ireland, September 2010.

\bibitem{hassani2010scaling1}
S.~H. Hassani and R.~Urbanke, ``{On the scaling of polar codes: I. The behavior
  of polarized channels},'' in \emph{Proc. IEEE International Symposium on
  Information Theory (ISIT)}, Austin, Texas, June 2010, pp. 874--878.

\bibitem{hassani2013rate}
S.~H. Hasani, R.~Mori, T.~Tanaka, and U.~R. {L}., ``{Rate-dependent analysis of
  the asymptotic behaviour of channel polarization},'' \emph{IEEE Transactions
  on Information Theory}, vol.~59, no.~4, pp. 2267--2276, April 2013.

\bibitem{korada2010empirical}
S.~B. Korada, A.~Montanari, E.~Telatar, and R.~Urbanke, ``{An empirical scaling
  law for polar codes},'' in \emph{Proc. IEEE International Symposium on
  Information Theory (ISIT)}, Austin, Texas, June 2010, pp. 884--888.

\bibitem{goli2012universal}
A.~Goli, S.~H. Hassani, and R.~Urbanke, ``{Universal bounds on the scaling
  behavior of polar codes},'' in \emph{Proc. IEEE International Symposium on
  Information Theory (ISIT)}, Boston, MA, July 2012, pp. 1957--1961.

\bibitem{hassani2013finite}
S.~H. Hassani, K.~Alishahi, and R.~Urbanke, ``{Finite-Length Scaling of polar
  codes},'' \emph{arXiv preprint arXiv:1304.4778}, 2013.

\bibitem{hassani2010scaling2}
------, ``{On the scaling of polar codes: II. The behavior of un-polarized
  channels},'' in \emph{Proc. IEEE International Symposium on Information
  Theory (ISIT)}, Austin, Texas, June 2010, pp. 879--883.

\bibitem{guruswami2013speed}
V.~Guruswami and P.~Xia, ``{Polar Codes: Speed of polarization and polynomial
  gap to capacity},'' \emph{arXiv preprint arXiv:1304.4321}, 2013.

\bibitem{korada2009polar}
S.~B. Korada, ``{Polar codes for channel and source coding},'' Ph.D.
  dissertation, EPFL, Lausanne, Switzerland, 2009.

\bibitem{ru_book}
T.~Richardson and R.~Urbanke, \emph{{Modern Coding Theory}}.\hskip 1em plus
  0.5em minus 0.4em\relax Cambridge, UK: Cambridge University Press, 2008.

\bibitem{khandekar2002graph}
A.~Khandekar, ``Graph-based codes and iterative decoding,'' Ph.D. dissertation,
  Citeseer, 2002.

\bibitem{karzand2010qsrc}
M.~Karzand and E.~Telatar, ``{Polar codes for q-ary source coding},'' in
  \emph{Proc. IEEE International Symposium on Information Theory (ISIT)},
  Austin, Texas, June 2010, pp. 909--912.

\bibitem{zhang1997redundancy}
Z.~Zhang, E.~Yang, and V.~K. Wei, ``{The redundancy of source coding with a
  fidelity criterion -- part 1: known statistics},'' \emph{IEEE Transactions on
  Information Theory}, vol.~43, no.~1, pp. 71--91, January 1997.

\bibitem{bur_bp_bounds}
D.~Burshtein and G.~Miller, ``{Bounds on the performance of belief propagation
  decoding},'' \emph{IEEE Transactions on Information Theory}, vol.~48, no.~1,
  pp. 112--122, Jan. 2002.

\end{thebibliography}

\end{document}